\documentclass[aps,pra,10pt,letterpaper,twocolumn,showpacs,superscriptaddress,floatfix]{revtex4-1}

\usepackage[colorlinks,linkcolor=blue,citecolor=red]{hyperref}
\usepackage{graphicx,epsfig}
\usepackage{amssymb,amsmath}

\newcommand{\rmB}{\mathrm{B}}
\newcommand{\rmC}{\mathrm{C}}
\newcommand{\rmL}{\mathrm{L}}
\newcommand{\rmM}{\mathrm{M}}
\newcommand{\rmO}{\mathrm{O}}
\newcommand{\rmR}{\mathrm{R}}
\newcommand{\rmI}{\mathrm{I}}
\newcommand{\rms}{\mathrm{s}}

\newcommand{\rmin}{\mathrm{in}}
\newcommand{\rmout}{\mathrm{out}}

\newcommand{\bfk}{\mathbf{k}}

\newcommand{\A}{\mathcal{A}}
\newcommand{\B}{\mathcal{B}}
\newcommand{\C}{\mathcal{C}}
\newcommand{\D}{\mathcal{D}}
\newcommand{\E}{\mathcal{E}}
\newcommand{\F}{\mathcal{F}}
\newcommand{\G}{\mathcal{G}}
\newcommand{\I}{\mathcal{I}}
\newcommand{\J}{\mathcal{J}}
\newcommand{\K}{\mathcal{K}}
\newcommand{\M}{\mathcal{M}}
\newcommand{\N}{\mathcal{N}}
\newcommand{\cP}{\mathcal{P}}
\newcommand{\Q}{\mathcal{Q}}
\newcommand{\R}{\mathcal{R}}
\newcommand{\T}{\mathcal{T}}
\newcommand{\V}{\mathcal{V}}
\newcommand{\Z}{\mathcal{Z}}

\newcommand{\trp}{\mathsf{T}}

\begin{document}

\title{Continuous variable entanglement swapping and its local certification: entangling distant mechanical modes}

\author{Mehdi Abdi}
\affiliation{Department of Physics, Iran University of Science and Technology, Tehran, Iran}
\author{Stefano Pirandola}
\affiliation{Department of Computer Science, University of York, York, United Kingdom}
\author{Paolo Tombesi}
\author{David Vitali}
\affiliation{School of Science and Technology, Physics Division, University of Camerino,
Camerino, Italy}

\date{\today}

\begin{abstract}
We introduce a modification of the standard entanglement swapping
protocol where the generation of entanglement between two distant
modes is realized and verified using only local optical
measurements. We show, indeed, that a simple condition on the
purity of the initial state involving also an ancillary mode is
sufficient to guarantee the success of the protocol by local
measurements
[\href{http://dx.doi.org/10.1103/PhysRevLett.109.143601}{M. Abdi
\textit{et al.}, Phys. Rev. Lett. \textbf{109}, 143601 (2012)}].
We apply the proposed protocol to a tripartite optomechanical
system where the never interacting mechanical modes become
entangled and certified using only local optical measurements.

\end{abstract}

\maketitle
%
%
\section{Introduction}
In quantum information networks, entanglement is a key feature for
secure exchange of information
\cite{Horodecki2009,Weedbrook2012,Braunstein2005,Nielsen2000}.
There are many proposals and realizations for generating
entanglement between various nodes of a quantum network;
entanglement of two trapped ions \cite{Turchette1998}, two atoms
\cite{Jaksch1999,Zheng2000}, two macroscopic diamonds at room
temperature~\cite{Walmsley2011} just to quote a few of them, up to
the most recent distribution of entanglement between distant
sites, as across a lake~\cite{Pan2012} or between two
islands~\cite{Zeilinger2012}. However, most proposals require
preparation through a physical, direct~\cite{Vitali2007} or
indirect~\cite{Cirac1997}, interaction. Entanglement swapping,
instead, is one of the most surprising effects of the non-locality
of quantum mechanics because it is a way to create entanglement,
i.e., quantum correlations,  between distant parties that never
interacted~\cite{Pan1998}. For continuous variables, which we are
here interested in, this technique was experimentally demonstrated
in Refs.~\cite{Jia2004,Takei2005}.

For nontrivial quantum communication tasks such as
teleportation~\cite{Pirandola2006tele,Weedbrook2012}, it is
necessary to ensure that the remote sites which are the ends of
the quantum channel are entangled. This condition may lead to
serious difficulties, since it requires test measurements on the
remote sites, which could be difficult to perform. Therefore, it
is important to test the success of a swapping protocol in easier
ways. In this paper we provide a solution to such a requirement,
i.e., we propose a protocol which makes it possible to test the
entanglement between remote nodes employing local optical
measurements only. Although our protocol imposes a condition on
the initially prepared states, from a practical point of view this
is a reasonable cost to pay for. Such a protocol can be utilized
for producing confident quantum channels between two far and
non-interacting nodes, e.g., two satellites, by measurements in
halfway for both creating and testing it (cf.
Fig.~\ref{fig:scheme}(b)). Moreover, this protocol provides a
promising method for experimentally creating entanglement between
two macroscopic objects in direction of questioning the so called
Schrodinger cat states and their
decoherence~\cite{Julsgaard2001,Romero-Isart2011a,Romero-Isart2011}.
From this point of view, this work extends
Ref.~\cite{Pirandola2006}, which first pioneered the possibility
to use entanglement swapping for entangling two massive systems,
such as two micromechanical oscillators.

In this paper the matrices are shown by curly capital letters, while the vectors are in bold face letters.
The paper is organized as follows:
In Sec.~II we explain the protocol. In Sec.~III we discuss the output state resulting from running the protocol.
Then, the protocol is applied in the specific case of optomechanical systems in Sec.~IV.
Concluding remarks are provided in Sec.~V.

%
%
\section{The protocol}
\begin{figure}[b]
\includegraphics[width=\columnwidth]{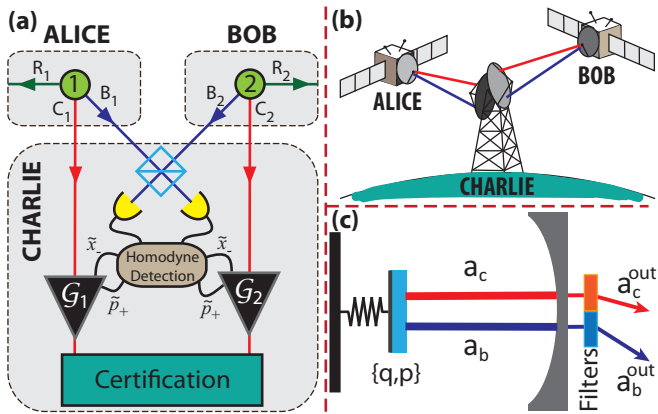}
\caption{(Color online) (a) Scheme of the entanglement swapping protocol with local certification. (b) Schematic quantum communication scenario in which the generalized entanglement swapping protocol applies. (c) The optomechanical setup which can be mounted on each remote site.}\label{fig:scheme}
\end{figure}

The initial states employed in this protocol, on both sides of the system, are tripartite continuous variable states.
In fact, the standard entanglement swapping protocol is modified by adding an ancillary mode to each side in order to provide the local certification of the achieved entanglement between the two remote sites.
In Fig.~\ref{fig:scheme}(a) the principles of the protocol is sketched.

\subsection{Initial state}
The whole system is initially composed of a pair of independent tripartite entangled bosonic modes, one possessed by Alice and one by Bob.
Alice and Bob are located at remote sites, prepare a \textit{specific} tripartite state, and each shares two modes with Charlie, who is located for simplicity halfway between them (see Fig.~\ref{fig:scheme}(a) and (b)).
These four bosonic modes (two modes from each side) will be manipulated by Charlie when executing the protocol.
Since the two initial subsystems of Alice and Bob are independent and non-interacting, the initial state of the whole system can be described by the tensor product of the two initial tripartite states, $\rho_{1}\otimes\rho_{2}$ where $\rho_{1}$ and $\rho_{2}$ are shared by Alice--Charlie and Bob--Charlie, respectively.
We identify the modes remained at each side (the `remote' modes) by the bosonic annihilation operators $\hat{a}_{l}$ with commutation relation $[\hat{a}_{l},\hat{a}_{l'}^{\dagger}]=\delta_{ll'}$.
The bosonic modes used for the Bell measurement and certifying process are described by $\hat{b}_{l}$, and $\hat{c}_{l}$ with similar commutators, respectively, where $l=1$ stands for Alice, while $l=2$ is related to Bob.
In the case of an optomechanical system which we will consider later on, the remote mode will be represented by a mechanical mode of a micro-mirror and the modes shared with Charlie by two output modes of the optical cavity (see Fig.~\ref{fig:scheme}(c)).

The tripartite initial state at each site must be prepared in a proper way, in order to achieve the desired state at the end of the swapping protocol.
That is, the final resulting state should give an entangled state between the remote modes and provide a trustful way to endorse it by the two ancillary, certifying, modes. This is satisfied when, in the output state, the remote modes are more entangled than the certifying modes, $E_{N}^{\rmR}>E_{N}^{\rmC}$, where $E_{N}$ is an entanglement monotone that in this paper is chosen to be the logarithmic negativity \cite{Vidal2002}.
In the following we shall derive an explicit condition for these \textit{certifying} tripartite states~\cite{Abdi2012}.

In the Wigner function formalism, the initial state is expressed by the product of the Wigner function of the states at each site,
\begin{equation}
W_{\rmin}(\alpha_{1},\alpha_{2},\beta_{1},\beta_{2},\gamma_{1},\gamma_{2}) =W_{1}(\alpha_{1},\beta_{1},\gamma_{1})W_{2}(\alpha_{2},\beta_{2},\gamma_{2}),
\end{equation}
where we have introduced the complex phase space variable $\alpha_{l} \equiv (x_{al}+ip_{al})/\sqrt{2}$ corresponding to the bosonic mode operator $\hat{a}_{l}$ and the same is done for the other modes, $\hat{b}_{l}\leftrightarrow\beta_{l}$ and $\hat{c}_{l}\leftrightarrow\gamma_{l}$ with $l=1,2$.
The real phase space variables $x_{kl}$ and $p_{kl}$ with $k=a,b,c$ and $l=1,2$ are the counterparts of the Hermitian quadrature operators $\hat{x}_{lk}$ and $\hat{p}_{kl}$ satisfying the commutation relations $[\hat{x}_{kl},\hat{p}_{k'l'}]=i\delta_{kk'}\delta_{ll'}$.

\subsection{Bell measurement}
In order to convert Alice--Charlie and Bob--Charlie entanglement to the nonlocal Alice--Bob entanglement, Charlie must erase some of the information shared with Alice and Bob. This is obtained via the CV version of the Bell measurement, which we recall here.

\textit{Beam-splitter mixing---}
Charlie mixes one mode from each side (here labeled as $\hat{b}_{1}$ and $\hat{b}_{2}$)
through a balanced beam-splitter, performing the following linear transformation for the annihilation operators
\begin{equation}
\left[\begin{array}{c} \hat{b}_{+} \\ \hat{b}_{-} \end{array}\right] =
\frac{1}{\sqrt{2}}\left[\begin{array}{cc}
                        1 & 1 \\
                        -1 & 1
\end{array}\right]
\left[\begin{array}{c} \hat{b}_{1} \\ \hat{b}_{2} \end{array}\right],
\end{equation}
where $\hat{b}_{\pm}$ refer to the output modes $\pm$ of the beam-splitter. At the level of quadratures one has
\begin{equation}
\hat{x}_{\pm}=\frac{\hat{x}_{b2} \pm \hat{x}_{b1}}{\sqrt{2}},~~
\hat{p}_{\pm}=\frac{\hat{p}_{b2} \pm \hat{p}_{b1}}{\sqrt{2}},
\end{equation}
while the phase space counterpart of this bilinear transformation is
$$\beta_{\pm}=\frac{\beta_{2}\pm\beta_{1}}{\sqrt{2}},$$
where $\beta_{\pm}\equiv (x_{\pm}+ip_{\pm})/\sqrt{2}$ are the complex phase-space variables associated with the beam-splitter output variables.

\textit{Homodyne detection.---}
Charlie measures the $\hat{x}_{-}$ and $\hat{p}_{+}$ quadratures with two homodyne detectors, getting the outcomes $\{\tilde{x}_{-},\tilde{p}_{+}\} =\sqrt{2}\{\tilde{\beta}_{-}^{\Re},\tilde{\beta}_{+}^{\Im}\}$ with probability $P(\tilde{x}_{-},\tilde{p}_{+})$, where the superscripts $\Re$ and $\Im$ correspond to the real and imaginary part of the complex number or variable.
This measurement leads to the conditional collapse of the initial six-mode state into a four-mode state:
\begin{equation}
W_{\rmin} \longrightarrow P(\tilde{x}_{-},\tilde{p}_{+})^{-1} W_{\rmin}
\delta(\beta_{-}^{\Re}-\tilde{\beta}_{-}^{\Re}) \delta(\beta_{+}^{\Im}-\tilde{\beta}_{+}^{\Im}).
\end{equation}
The conditional output state generated at this stage is obtained by tracing out the beam-splitter output modes $\pm$, i.e., by integrating the collapsed state Wigner function over the variables $\beta_{+}$ and $\beta_{-}$, which reads
\begin{align}\label{wigcon0}
&W_{\mathrm{con}}(\alpha_{1},\alpha_{2},\gamma_{1},\gamma_{2}|\tilde{\beta}_{-}^{\Re},\tilde{\beta}_{+}^{\Im})=\frac{1}{P(\tilde{\beta}_{-}^{\Re},\tilde{\beta}_{+}^{\Im})} \nonumber \\
&\times\int d\beta_{-}^{\Im} \int d\beta_{+}^{\Re} W_{\rmin}(\alpha_{1},\alpha_{2},\gamma_{1},\gamma_{2},\beta_{+},\beta_{-})|_{{\beta}_{-}^{\Re}=\tilde{\beta}_{-}^{\Re},{\beta}_{+}^{\Im}=\tilde{\beta}_{+}^{\Im}}.
\end{align}
By introducing $\tilde{\beta} \equiv i\tilde{p}_{+} -\tilde{x}_{-} =\sqrt{2}(i\tilde{\beta}_{+}^{\Im} -\tilde{\beta}_{-}^{\Re})$, which is a complex number representing the measurement outcomes in a compact form, and $\beta \equiv [x_+ +i\tilde{p}_+ -(\tilde{x}_- +ip_-)]/2$ which is actually equal to $\beta_{1}|_{\{\tilde{x}_-,\tilde{p}_+\}}$ (i.e., $\beta_1$ specified by the measurement outcomes) we arrive at the following compact form for the conditional state after the Bell measurement
\begin{align}\label{wigcon}
W_{\mathrm{con}}(\alpha_{1},\alpha_{2},\gamma_{1},\gamma_{2}|\tilde{\beta})=&\frac{1}{P(\tilde{\beta})}\int d^{2}\beta W_{1}(\alpha_{1},\gamma_{1},\beta) \nonumber \\
&~~~~~\times W_{2}(\alpha_{2},\gamma_{2},\beta^{*}-\tilde{\beta}^{*}).
\end{align}
Eq.~(\ref{wigcon}) has been obtained using the fact that $\beta_{2}|_{\{\tilde{x}_{-},\tilde{p}_{+}\}} =\beta^*-\tilde{\beta}^*$,
the property
\begin{equation}
P(\tilde{\beta}_{-}^{\Re},\tilde{\beta}_{+}^{\Im})=P(\tilde{\beta}_{-}^{\Re}|\tilde{\beta}_{+}^{\Im})P(\tilde{\beta}_{+}^{\Im}),
\end{equation}
and that $\tilde{\beta}^{\Re}=-\sqrt{2}\tilde{\beta}_{-}^{\Re}$ and $\tilde{\beta}^{\Im}=\sqrt{2}\tilde{\beta}_{+}^{\Im}$.
Moreover, we have also exploited the fact that $P(ky)=P(y)/|k|$ for $k\in \mathbb{R}$, so that
\begin{align}
P(\tilde{\beta}_{-}^{\Re},\tilde{\beta}_{+}^{\Im}) &=\sqrt{2}P(\tilde{\beta}^{\Re}|\tilde{\beta}^{\Im}) \sqrt{2}P(\tilde{\beta}^{\Im}) =2P(\tilde{\beta}^{\Re},\tilde{\beta}^{\Im}) \nonumber \\
&\equiv 2P(\tilde{\beta}),
\end{align}
and that $\int d\beta_{-}^{\Im}\int d\beta_{+}^{\Re} \leftrightarrow 2\int d^{2}\beta$.

\subsection{Classical communication}
The conditional state of Eq.~(\ref{wigcon}) has a fluctuating displacement associated with the outcome of the Bell measurement. Charlie broadcasts the measurement results, so that Charlie himself, as well as Alice and Bob, may suitably displace their modes according to the measurement outcomes.
In the Heisenberg picture, these displacements, which will complete the swapping process, are \cite{Loock1999}
\begin{subequations}
\begin{eqnarray}
&&\bigg\{\begin{array}{ll}
\hat{x}_{a1} \rightarrow \hat{x}_{a1} +\sqrt{2} ~\tilde{x}_{-} \\
\hat{p}_{a1} \rightarrow \hat{p}_{a1} +\sqrt{2} ~\tilde{p}_{+}
\end{array}, \\
&&\bigg\{\begin{array}{ll}
\hat{x}_{a2} \rightarrow \hat{x}_{a2} -\sqrt{2} ~\tilde{x}_{-} \\
\hat{p}_{a2} \rightarrow \hat{p}_{a2} +\sqrt{2} ~\tilde{p}_{+}
\end{array}, \\
&&\bigg\{\begin{array}{ll}
\hat{x}_{c1} \rightarrow \hat{x}_{c1} +\sqrt{2} ~\tilde{x}_{-} \\
\hat{p}_{c1} \rightarrow \hat{p}_{c1} +\sqrt{2} ~\tilde{p}_{+}
\end{array}, \\
&&\bigg\{\begin{array}{ll}
\hat{x}_{c2} \rightarrow \hat{x}_{c2} -\sqrt{2} ~\tilde{x}_{-} \\
\hat{p}_{c2} \rightarrow \hat{p}_{c2} +\sqrt{2} ~\tilde{p}_{+}
\end{array}.
\end{eqnarray}
\end{subequations}
However, in practice, Alice, Bob, and Charlie may employ gain factors in displacing their modes~\cite{Hoelscher-Obermaier2011}.
As it will be discussed in Sec.~III, application of these gain factors may be useful for improving the quality of the swapped entanglement.
In terms of the complex phase space variables, these conditional displacements can be expressed as
\begin{subequations}
\label{complex}
\begin{eqnarray}
&&\alpha_{1} \rightarrow \alpha_{1} +\tilde{\beta}_{a1}^{*}, \\
&&\alpha_{2} \rightarrow \alpha_{2} -\tilde{\beta}_{a2}, \\
&&\gamma_{1} \rightarrow \gamma_{1} +\tilde{\beta}_{c1}^{*}, \\
&&\gamma_{2} \rightarrow \gamma_{2} -\tilde{\beta}_{c2},
\end{eqnarray}
\end{subequations}
where the displacement of each mode when phase-sensitive gain factors are used are given by
\begin{subequations}
\label{displacements}
\begin{eqnarray}
\tilde{\beta}_{a1} &=&-g_{a1}^{\Re} \tilde{x}_{-} +i g_{a1}^{\Im} \tilde{p}_{+}, \\
\tilde{\beta}_{a2} &=&-g_{a2}^{\Re} \tilde{x}_{-} +i g_{a2}^{\Im} \tilde{p}_{+}, \\
\tilde{\beta}_{c1} &=&-g_{c1}^{\Re} \tilde{x}_{-} +i g_{c1}^{\Im} \tilde{p}_{+}, \\
\tilde{\beta}_{c2} &=&-g_{c2}^{\Re} \tilde{x}_{-} +i g_{c2}^{\Im} \tilde{p}_{+}.
\end{eqnarray}
\end{subequations}
In practice, the process is run continuously with measurement
outcomes changing in time, so that the conditional state
$W_{\mathrm{con}}$ of Eq.~(\ref{wigcon}) is transformed into a
displaced state $W_{\mathrm{dis}}$ according to
Eqs.~(\ref{complex}) with probability $P(\tilde{\beta})$. In
general, the state of the system is therefore given by the
ensemble average
\begin{equation}\label{wigens}
W_{\mathrm{ens}}(\alpha_{1},\alpha_{2},\gamma_{1},\gamma_{2})=\int
d^2\tilde{\beta}P(\tilde{\beta})W_{\mathrm{dis}}(\alpha_{1},\alpha_{2},\gamma_{1},\gamma_{2}|\tilde{\beta}).
\end{equation}
We remark that this average is superfluous if the displacements
are optimal, such to transform $W_{\mathrm{con}}$ into a zero-mean
state \cite{Pirandola2006}. As we will see afterwards, this
reduction is also exploited in our approach.

%
%
\section{The output state}
The output state of the swapping protocol is described by Eq.~(\ref{wigens}) which completely characterizes the final state of the system and is given by a convolution integral of the Wigner functions of the factorized initial state, evaluated at appropriate phase space points. For this reason it is convenient to express the output state in terms
of its symmetrically-ordered characteristic function which is just the Fourier transform of the Wigner function, $\Phi(\lambda_{1},\lambda_{2},\mu_{1},\mu_{2})=\mathrm{FT}[W(\alpha_{1},\alpha_{2},\gamma_{1},\gamma_{2})]$, obtaining
\begin{equation}\label{chf}
\Phi_{\mathrm{ens}}(\lambda_{1},\lambda_{2},\mu_{1},\mu_{2})=\Phi_{1}(\lambda_{1},\mu_{1},\nu)\Phi_{2}(\lambda_{2},\mu_{2},\nu^*),
\end{equation}
where $\lambda_{k}$ and $\mu_{k}$ are the conjugate variables for $\alpha_{k}$ and $\gamma_{k}$ in the Wigner function, while the correlations between the four modes are contained in the complex variable $\nu$ given by
\begin{align}\label{nu}
\nu &\equiv g_{a1}^{\Im}\lambda_{1}^{\Re} +g_{c1}^{\Im}\mu_{1}^{\Re}
+g_{a2}^{\Im}\lambda_{2}^{\Re} +g_{c2}^{\Im}\mu_{2}^{\Re} \nonumber \\
&~~~+i(g_{a2}^{\Re}\lambda_{2}^{\Im} +g_{c2}^{\Re}\mu_{2}^{\Im} -g_{a1}^{\Re}\lambda_{1}^{\Im} -g_{c1}^{\Re}\mu_{1}^{\Im}).
\end{align}
In order to perform calculations, it is convenient to adopt a vector notation in which we associate to each complex variable a two-dimensional real vector according to
\begin{equation}
\lambda=\lambda^{\Re}+i\lambda^{\Im} \longleftrightarrow \boldsymbol{\lambda}\equiv [\lambda^{\Im},-\lambda^{\Re}]^{\trp}.
\end{equation}
As a consequence, the characteristic functions in Eq.~(\ref{chf}) can be rewritten as
\begin{widetext}
\begin{align}
\Phi_{1}(\lambda_{1},\mu_{1},\nu)&\longleftrightarrow\Phi_{1}(\boldsymbol{\lambda}_{1},\boldsymbol{\mu}_{1},\G_{a1}\boldsymbol{\lambda}_{1} +\G_{a2}\boldsymbol{\lambda}_{2} +\G_{c1}\boldsymbol{\mu}_{1} +\G_{c2}\boldsymbol{\mu}_{2}), \\
\Phi_{2}(\lambda_{2},\mu_{2},\nu^{*})&\longleftrightarrow\Phi_{2}(\boldsymbol{\lambda}_{2},\boldsymbol{\mu}_{2},-\Z\G_{a1}\boldsymbol{\lambda}_{1} -\Z\G_{a2}\boldsymbol{\lambda}_{2} -\Z\G_{c1}\boldsymbol{\mu}_{1} -\Z\G_{c2}\boldsymbol{\mu}_{2}),
\end{align}
\end{widetext}
where $\Z\equiv\mathrm{diag}[1,-1]$ and we have introduced the following gain matrices
\begin{align}
\G_{a1}&\equiv \left[\begin{array}{cc} -g_{a1} & 0 \\ 0 & h_{a1} \end{array}\right],~~
\G_{a2}\equiv \left[\begin{array}{cc} g_{a2} & 0 \\ 0 & h_{a2} \end{array}\right], \\
\G_{c1}&\equiv \left[\begin{array}{cc} -g_{c1} & 0 \\ 0 & h_{c1} \end{array}\right],~~
\G_{c2}\equiv \left[\begin{array}{cc} g_{c2} & 0 \\ 0 & h_{c2} \end{array}\right].
\end{align}
%

\subsection{The case of initial tripartite Gaussian states}
We now restrict to the physically relevant case when the two
independent tripartite states $\rho_1$ and $\rho_2$ at Alice and
Bob sites are Gaussian. For the class of Gaussian states, the
characteristic function is completely determined by the first and
second moments of the quadrature operators \cite{Weedbrook2012}.
In fact, for an $N$-mode Gaussian state, the characteristic
function is equal to $\Phi(\bfk)=\exp\{-\bfk^{\trp} \V \bfk/2 +i
\mathbf{d}^{\trp}\bfk\}$, where $\V$ and $\mathbf{d}$ are the
covariance matrix (CM) and displacement vector of the state,
respectively, and $\bfk=[x_{1},p_{1},...,x_{N},p_{N}]^{\trp}$ is
the vector of phase space variables. The entanglement properties
of the final state are fully determined by the CM because the
displacement affects only local properties.

We consider two initial tripartite Gaussian states with zero displacement and characterized by the following CM
\begin{equation}\label{incm}
\V_{\mathrm{k}} =\left[\begin{array}{ccc}
    \R_k & \D_k & \F_k \\
    \D_k^{\trp} & \B_k & \E_k \\
    \F_k^{\trp} & \E_k^{\trp} & \C_k
    \end{array}\right], \;\; k =1,2,
\end{equation}
which is expressed in terms of its $2 \times 2$ sub-blocks. By inserting the corresponding characteristic functions into Eq.~(\ref{chf}), one gets for the ensemble-averaged output state a four-mode Gaussian state with vanishing first moments and a CM given by
\begin{equation}\label{enscm}
\V_{\rmin} =\left[\begin{array}{cc}
    \V_{\mathrm{1}} &  \\
     & \V_{\mathrm{2}}
  \end{array}\right] \longrightarrow
\V_{\mathrm{ens}} =\left[\begin{array}{cc}
    \V_{\rmR}' & \V_{\mathrm{X}}' \\
    \V_{\mathrm{X}}^{\prime\trp} & \V_{\rmC}'
  \end{array}\right].
\end{equation}
In particular, the CM of the interesting bipartite subsystems (Alice--Bob and the certifying modes) are given by
\begin{widetext}
\begin{align}
\V_{\rmR}'&=\left[ \begin{array}{cc}
                \R_1 &  \\
                    & \R_2
                \end{array} \right]
                +\left[ \begin{array}{cc}
        \G_{a1}^{\trp}\M\G_{a1} +\D_1^{\trp}\G_{a1} +\G_{a1}^{\trp}\D_1 & \G_{a1}^{\trp}\M\G_{a2} +\D_1^{\trp}\G_{a2} -\G_{a1}^{\trp}\Z\D_2 \\
        \G_{a2}^{\trp}\M\G_{a1} +\G_{a2}^{\trp}\D_1 -\D_2^{\trp}\Z\G_{a1} & \G_{a2}^{\trp}\M\G_{a2} -\D_2^{\trp}\Z\G_{a2} +\G_{a2}^{\trp}\Z\D_2
                \end{array} \right], \\
\V_{\rmC}'&=\left[ \begin{array}{cc}
                \C_1 &  \\
                    & \C_2
                \end{array} \right]
                +\left[ \begin{array}{cc}
        \G_{c1}^{\trp}\M\G_{c1} +\E_1^{\trp}\G_{c1} +\G_{c1}^{\trp}\E_1 & \G_{c1}^{\trp}\M\G_{c2} +\E_1^{\trp}\G_{c2} -\G_{c1}^{\trp}\Z\E_2 \\
        \G_{c2}^{\trp}\M\G_{c1} +\G_{c2}^{\trp}\E_1 -\E_2^{\trp}\Z\G_{c1} & \G_{c2}^{\trp}\M\G_{c2} -\E_2^{\trp}\Z\G_{c2} +\G_{c2}^{\trp}\Z\E_2
                \end{array} \right],
\end{align}
\end{widetext}
where we have introduced the matrix $\M \equiv \B_1+\Z\B_2\Z$.

\subsection{Optimization of the output state}
The ensemble average output state is of much less quality and less
entangled than the initial state because of the average over the
differently displaced states conditioned to the homodyne
measurement outcome. However one can optimize the output state by
optimizing the choice of the gain factors. It is quite evident
that such an optimization corresponds to adjust the gain so that
the displacement of the conditional state is always put to zero.
In such a case the output state is no more blurred by the
fluctuating measurement outcomes and the CM of the output state
corresponds to that of the conditional state \cite{Pirandola2006}.

The first moment of the displaced conditional state can be obtained by calculating the characteristic function of the Wigner function which is obtained from Eq.~(\ref{wigcon})
\begin{align}
W_{\mathrm{dis}}=\frac{1}{P(\tilde{\beta})}&\int d^{2}\beta W_{1}(\alpha_{1}+\tilde{\beta}_{a1}^{*},\gamma_{1}+\tilde{\beta}_{c1}^{*},\beta) \nonumber \\
&\times W_{2}(\alpha_{2}-\tilde{\beta}_{a2},\gamma_{2}-\tilde{\beta}_{c2},\beta^{*}-\tilde{\beta}^{*}),
\end{align}
which is given by
\begin{align}
\Phi_{\mathrm{dis}}(\lambda_{1},\lambda_{2},\mu_{1},\mu_{2})&= \frac{1}{\pi^{2}P(\tilde{\beta})}\int d^{2}\eta \Phi_{1}(\lambda_{1},\mu_{1},\eta^{*}) \nonumber \\ &\times\exp\{-\mu_{1}\tilde{\beta}_{c1}-\mu_{1}^{*}\tilde{\beta}_{c1}^{*}\} \Phi_{2}(\lambda_{2},\mu_{2},\eta) \nonumber \\
&\times \exp\{\mu_{2}\tilde{\beta}_{c2}^{*}-\mu_{2}^{*}\tilde{\beta}_{c2}\}.
\end{align}
Now let us switch to the vector notation, by defining the vector
corresponding to the measurement outcome
$\tilde{\boldsymbol{\beta}}=[i\tilde{\beta}^{\rmR},-i\tilde{\beta}^{\rmI}]$,
so that this characteristic function can be rewritten as
\begin{eqnarray}
&&\Phi_{\mathrm{dis}}(\boldsymbol{\lambda_1},\boldsymbol{\lambda_2},\boldsymbol{\mu_1},\boldsymbol{\mu_2})=\frac{\exp\Big\{2\tilde{\boldsymbol{\beta}}^{\trp}(\Z\G_{c1}\boldsymbol{\mu}_{1}+\Z\G_{c2}\boldsymbol{\mu}_{2})\Big\} }{\pi^{2} P(\tilde{\boldsymbol{\beta}})} \nonumber \\
&&\times\int d^{2}\boldsymbol{\eta} \exp\{2\tilde{\boldsymbol{\beta}}^{\trp}\boldsymbol{\eta}\} \Phi_{1}(\boldsymbol{\lambda}_{1},\boldsymbol{\mu}_{1},-\Z\boldsymbol{\eta}) \Phi_{2}(\boldsymbol{\lambda}_{2},\boldsymbol{\mu}_{2},\boldsymbol{\eta}).
\end{eqnarray}
Since we have considered the initial state of each side to be a
zero-displaced Gaussian state, whose CM is given by
Eq.~(\ref{incm}), we arrive at the following first moment vector
for the displaced conditional state
\begin{equation}\label{dvector}
\mathbf{d}_{\mathrm{dis}}=-2i
\left[\begin{array}{c}
\G_{a1}\Z +\D_1 \Z\M^{-1})\tilde{\boldsymbol{\beta}}\\
(\G_{c1}\Z +\E_1 \Z\M^{-1})\tilde{\boldsymbol{\beta}}\\
(\G_{c2}\Z -\E_2 \M^{-1})\tilde{\boldsymbol{\beta}}\\
(\G_{a2}\Z -\D_2 \M^{-1})\tilde{\boldsymbol{\beta}}
\end{array}\right].
\end{equation}
By applying the condition for the optimal output state, i.e. $\mathbf{d}_{\mathrm{dis}}=\mathbf{0}$, from Eq.~(\ref{dvector}) we get
\begin{subequations}
\begin{eqnarray}
\G_{a1} &=&-\Z\M^{-1}\Z\D_1, \\
\G_{a2} &=&\Z\M^{-1}\D_2, \\
\G_{c1} &=&-\Z\M^{-1}\Z\E_1, \\
\G_{c2} &=&\Z\M^{-1}\E_2,
\end{eqnarray}
\end{subequations}
as the optimal values for the gain matrices. Finally, the CM of
the optimally displaced (output) state reads
\begin{equation}\label{outcm}
\V_{\mathrm{out}} =\left[\begin{array}{cc}
    \V_{\rmR} & \V_{\mathrm{X}} \\
    \V_{\mathrm{X}}^{\trp} & \V_{\rmC}
  \end{array}\right],
\end{equation}
which is identical to the CM of the conditional state, expressed
by the Wigner function in Eq.~(\ref{wigcon}). Explicitly, the
various blocks $\V_{\rmR}$, $\V_{\rmC}$, and $\V_{\mathrm{X}}$ are
equal to~\cite{Abdi2012}
\begin{align}
\V_{\rmR}&=\left[\begin{array}{cc}
            \R_1 -\D_1^{\trp}\Z\M^{-1}\Z\D_1 & \D_1^{\trp}\Z\M^{-1}\D_2 \\
            \D_2^{\trp}\M^{-1}\Z\D_1 & \R_2 -\D_2^{\trp}\M^{-1}\D_2
            \end{array}\right], \\
\V_{\rmC}&=\left[\begin{array}{cc}
            \C_1 -\E_1^{\trp}\Z\M^{-1}\Z\E_1 & \E_1^{\trp}\Z\M^{-1}\E_2 \\
            \E_2^{\trp}\M^{-1}\Z\E_1 & \C_2 -\E_2^{\trp}\M^{-1}\E_2
            \end{array}\right], \\
\V_{\mathrm{X}}&=\left[\begin{array}{cc}
            \F_1 -\D_1^{\trp}\Z\M^{-1}\Z\E_1 & \D_1^{\trp}\Z\M^{-1}\E_2 \\
            \D_2^{\trp}\M^{-1}\Z\E_1 & \F_2 -\D_2^{\trp}\M^{-1}\E_2
            \end{array}\right].
\end{align}
%

\subsection{Standard form}
To get an intuitive picture for determining the conditions under which the entanglement swapping with local certification protocol properly works we use the standard form of the CM.
In fact, the CM of an arbitrary $N$-mode state expresses the covariances between the quadratures of the state, and, for this reason, it must respect the uncertainty principle.
Therefore, we adopt the compact form of commutation relation for the vector of operators, $\hat{\bfk}=[\hat{x}_1,\hat{p}_1,...,\hat{x}_N,\hat{p}_N]^{\trp}$, as $[\hat{\bfk}_l,\hat{\bfk}_m]=i\J_{lm}^{(N)}$, where
\begin{equation}
\J^{(N)}=\bigoplus_{k=1}^N \J_k ~,~~~\mathrm{with}~~
\J_k \equiv \left[\begin{array}{cc}
             0 & 1 \\
            -1 & 0
            \end{array}\right],
\end{equation}
is the $N$-mode symplectic form. Thus, every CM must satisfy the
following condition
\begin{equation}\label{genuinecm}
\V+\frac{i}{2}\J^{(N)} \geq 0.
\end{equation}
The results of the two previous sections can be expressed in a
simplified way by exploiting the standard form of the CM. The CM
of every tripartite system can be transformed in the following
form via local unitary operators \cite{Adesso2006}
\begin{equation}\label{stdform}
\V =\left[\begin{array}{cccccc}
    r & 0 & d & 0 & f & f' \\
    0 & r & 0 & d' & f'' & f''' \\
    d & 0 & b & 0 & e & e'' \\
    0 & d' & 0 & b & 0 & e' \\
    f & f'' & e & 0 & c & 0 \\
    f' & f''' & e'' & e' & 0 & c
  \end{array}\right].
\end{equation}
Applying this standard form to the CMs of the initial tripartite states in Eq.~(\ref{incm}) is equivalent to set $\R_{k}=r_{k}\I$, $\B_{k}=b_{k}\I$, and $\C_{k}=c_{k}\I$ where $\I$ is the $2\times 2$ identity matrix.
Also we have $\D_{k}=\mathrm{diag}[d_{k},d'_{k}]$, and
\begin{equation}
\E_{k}=\left[\begin{array}{cc}
        e_{k} & e''_{k} \\
        0 & e'_{k}
        \end{array}\right], ~~
\F_{k}=\left[\begin{array}{cc}
        f_{k} & f'_{k} \\
        f''_{k} & f'''_{k}
        \end{array}\right], \nonumber
\end{equation}
where $k=1,2$. However, when all $2\times 2$ submatrices of the CM
$\V_{k}$ are non-singular, the standard form of
Eq.~(\ref{stdform}) gets an additional zero element $e''_{k}=0$,
i.e., we can write $\E_{k}=\mathrm{diag}[e_{k},e'_{k}]$ (cf.
Ref.~\cite{Wang2003}).

As an entanglement monotone, we adopt the logarithmic negativity
\cite{Vidal2002}
\begin{equation}\label{logneg}
E_{N}= \mathrm{max}\{0,-\ln 2\eta^{-}\},
\end{equation}
where $\eta^{-}$ is the minimum symplectic eigenvalue of the
partially transposed CM. This is also known as minimum
partially-transposed symplectic (PTS) eigenvalue and it is given
by
\begin{equation}
\eta^{-}=\frac{1}{\sqrt{2}}\Big(\Sigma(\V) - \sqrt{\Sigma(\V)^{2}
-4 \det \V}\Big)^{\frac{1}{2}},
\end{equation}
where $\Sigma(\V) \equiv \det\A + \det\B - 2\det\C$ can be
extracted from the original CM expressed in the block form
\begin{equation}
\V =\left[\begin{array}{cc} \A & \C \\ \C^{\trp} & \B \end{array}\right].
\end{equation}
It is clear that entanglement is present when $E_{N}>0$ or
equivalently $\eta^{-}<1/2$. Furthermore, $\eta^{-}$ is itself an
entanglement monotone for Gaussian states, since it is
monotonically related to $E_{N}$.

From the standard form of Eq.~(\ref{stdform}), we arrive at the
following bipartite CMs for the two bipartite subsystems
describing the two remote network nodes --Alice and Bob-- and the
certifying parties of Charlie,
\begin{widetext}
\begin{align}
\V_{\rmR} &=\frac{1}{b_{1}+b_{2}}
\left[\begin{array}{cccc}
      r_{1}(b_{1}+b_{2})-d_{1}^2 & 0 & d_{1}d_{2} & 0 \\
      0 & r_{1}(b_{1}+b_{2})-d'^{2}_{1} & 0 & -d'_{1}d'_{2} \\
      d_{1}d_{2} & 0 & r_{2}(b_{1}+b_{2})-d_{2}^{2} & 0 \\
      0 & -d'_{1}d'_{2} & 0 & r_{2}(b_{1}+b_{2})-d'^{2}_{2}
      \end{array}\right], \\
\V_{\rmC} &=\frac{1}{b_{1}+b_{2}}
\left[\begin{array}{cccc}
      c_{1}(b_{1}+b_{2})-e_{1}^{2} & -e_{1}e_{1}'' & e_{1}e_{2} & e_{1}e_{2}'' \\
      -e_{1}e_{1}'' & c_{1}(b_{1}+b_{2})-(e'^{2}_{1}+e''^{2}_{1}) & e_{1}''e_{2} & e''_{1}e_{2}''-e'_{1}e_{2}' \\
      e_{1}e_{2} & e_{1}''e_{2} & c_{2}(b_{1}+b_{2})-e_{2}^{2} & -e_{2}e''_{2} \\
      e_{1}e_{2}'' & e''_{1}e''_{2}-e'_{1}e'_{2} & -e_{2}e''_{2} & c_{2}(b_{1}+b_{2})-(e'^{2}_{2}+e''^{2}_{2})
      \end{array}\right].
\end{align}
\end{widetext}
From these matrices the remote--remote and certifying bipartite entanglement can be calculated, but their expression is too cumbersome to be reported here.

However, the explicit values for the entanglement monotone of the two bipartite states is significantly simplified when the initial tripartite states are identical, that is, when we start from a perfectly symmetric state between Alice and Bob.
In this simpler case, we are able to derive compact formulas for the partial transpose symplectic eigenvalues. In fact one gets for the CM of the remote modes
\begin{equation}
\V_{\rmR} = \left[
\begin{array}{cccc}
    r -\frac{d^{2}}{2b} &   & \frac{d^{2}}{2b} &   \\
      & r -\frac{d'^{2}}{2b} &   & -\frac{d'^{2}}{2b} \\
    \frac{d^2}{2b} &   & r -\frac{d^2}{2b} &   \\
      & -\frac{d'^{2}}{2b} &   & r -\frac{d'^{2}}{2b}
\end{array} \right],
\end{equation}
whose partial transpose CM $\V_{\rmR}^{\mathsf{PT}}$ has
symplectic eigenvalues $\eta_{\rmR}^{-}=b^{-1}\sqrt{\det
\V_{\mathrm{RB}}}$ and $\eta_{\rmR}^{+}=r$. On the other hand, the
CM corresponding to the Charlie's test parties takes the following
form
\begin{equation}
\V_{\rmC} = \left[
\begin{array}{cccc}
    c -\frac{e^{2}}{2b} & -\frac{ee''}{2b} & \frac{e^{2}}{2b} & \frac{ee''}{2b} \\
    -\frac{ee''}{2b} & c -\frac{e'^{2}+e''^{2}}{2b} & \frac{ee''}{2b} & \frac{e''^{2}-e'^{2}}{2b} \\
    \frac{e^{2}}{2b} & \frac{ee''}{2b} & c -\frac{e^{2}}{2b} & -\frac{ee''}{2b} \\
    \frac{ee''}{2b} & \frac{e''^{2}-e'^{2}}{2b} & -\frac{ee''}{2b} & c -\frac{e'^{2}+e''^{2}}{2b}
\end{array} \right],
\end{equation}
and the symplectic eigenvalues of the partial transpose CM
$\V_{\rmC}^{\mathsf{PT}}$ are $\eta_{\rmC}^{\pm}
=(b\sqrt{2})^{-1}\Big[\det\V_{\mathrm{BC}}+b^{2}c^{2} \pm \sqrt{(
b^{2}c^{2} - \det\V_{\mathrm{BC}})^{2} - ( 2bce'e'')^{2}}
\Big]^{\frac{1}{2}}$. In the above equations, $\V_{\mathrm{RB}}$
and $\V_{\mathrm{BC}}$ are the CM of the input subsystems given by
\begin{equation}
V_{\mathrm{RB}}= \left[
\begin{array}{cc}
    \R & \D   \\
    \D^{\trp} & \B
\end{array} \right], ~~~
V_{\mathrm{BC}}= \left[
\begin{array}{cc}
    \B & \E \\
    \E^{\trp} & \C
\end{array} \right].
\end{equation}
%

\subsection{Tripartite certifying states}
The condition for a successful, locally certified, entanglement
swapping is obtained by finding the relation between the
entanglement monotones $\eta_{R}^{-}$ and $\eta_{C}^{-}$ of the
two bipartite subsystems, the remote modes at Alice and Bob sites,
and the certifying modes in Charlie's hands. In the symmetric
case, such a relation is given by their ratio,
\begin{widetext}
\begin{equation}
\frac{\eta_{\rmC}^{-}}{\eta_{\rmR}^{-}} = \bigg[\frac{\det\V_{\mathrm{BC}}+b^{2}c^{2} - \sqrt{(b^{2}c^{2}-\det\V_{\mathrm{BC}})^{2} -(2bce'e'')^{2}}}{2\det\V_{\mathrm{RB}}}\bigg]^{-\frac{1}{2}}.
\end{equation}
\end{widetext}
If we consider the standard form of Ref.~\cite{Wang2003}, i.e. $e_{k}''=0$, the relation between the remote sites entanglement and the certifying entanglement takes the following general form
\begin{equation}
\eta_{\rmR}^{-} = \chi \eta_{\rmC}^{-},
\end{equation}
where $\chi$ is a local symplectic invariant given by
\begin{equation}\label{cert}
\chi = \sqrt{\frac{\det \V_{\mathrm{RB}}}{\det \V_{\mathrm{BC}}}}.
\end{equation}
This gives a sufficient condition for an indirectly measurable
entanglement between Alice and Bob. In other words, if any
entanglement between the certifying modes is detected by Charlie,
the two distant and non-interacting modes are surely entangled,
provided that the initial states are prepared so that $\chi<1$.
This sufficient condition can be expressed in terms of local
purities of the system. The purity of a state is defined as
$\mu(\varrho)=\mathrm{Tr}(\varrho^{2})$, where for an $N$-mode
Gaussian state with CM $\V(\varrho)$ is equal to
\begin{equation}
\mu(\varrho)=\frac{1}{2^{N}\sqrt{\det\V(\varrho)}}.
\end{equation}
Therefore, it is easy to show that the minimum PTS eigenvalues of
the bipartite remote sites and certifying modes are related to the
local purities by the following equations
\begin{equation}\label{purities}
\eta_{\rmR}^{-}=\frac{\mu_{\rmB}}{2\mu_{\rmR\rmB}}, ~~
\eta_{\rmC}^{-}=\frac{\mu_{\rmB}}{2\mu_{\rmB\rmC}},
\end{equation}
where $\mu_{\rmB}$ is purity of the Bell mode, $\mu_{\rmR\rmB}$ that of the system formed by the mode in the remote site and that subject to the Bell measurement, and $\mu_{\rmB\rmC}$ that of the system formed by the two modes at Charlie's site.
The sufficient and necessary condition for a successfully certified swapping process is to have for the output state $E_{N}^{\rmR}>E_{N}^{\rmC}>0$.
This condition can be rewritten from Eqs.~(\ref{logneg}) and (\ref{purities}) as
\begin{equation}\label{certp}
\mu_{\rmR\rmB} > \mu_{\rmB\rmC} > \mu_{\rmB}.
\end{equation}
Notice that this necessary and sufficient condition for ensuring
that the swapping process is successfully executed and certified
implies that the initial tripartite state should be prepared such
that the certifying--Bell and remote--Bell bipartite subsystems
are entangled. This can be verified as follows. According to
Refs.~\cite{Adesso2004,Adesso2004a} the bipartite Gaussian state
of the remote-Bell subsystem is inseparable if and only if
\begin{equation}
\mu_{\rmR\rmB}>\frac{\mu_{\rmR}\mu_{\rmB}}{\sqrt{\mu_{\rmR}^{2}+\mu_{\rmB}^{2}-\mu_{\rmR}^{2}\mu_{\rmB}^{2}}}.
\end{equation}
However, for every two variables $x$ and $y$ confined to $0\leq x,y \leq 1$, the inequality $$x\geq\frac{xy}{\sqrt{x^{2}+y^{2}-x^{2}y^{2}}}$$ is always true. Therefore, by setting $x=\mu_{\rmB}$ and $y=\mu_{\rmR}$ and using Eq.~(\ref{certp}), one has
\begin{equation}
\mu_{\rmR\rmB}>\mu_{\rmB}\geq \frac{\mu_{\rmR}\mu_{\rmB}}{\sqrt{\mu_{\rmR}^{2}+\mu_{\rmB}^{2}-\mu_{\rmR}^{2}\mu_{\rmB}^{2}}},
\end{equation}
which is just the necessary and sufficient condition for the entanglement of the remote--Bell subsystem.
The same argument can be applied for the certifying--Bell subsystem by putting  $x=\mu_{\rmB}$ and $y=\mu_{\rmC}$.

%
%
\section{The optomechanical system}
We now apply this protocol to the case of an optomechanical
system, in order to achieve entanglement between two distant
macroscopic mechanical resonators. To this end, one prepares a
tripartite optomechanical system involving a mechanical resonator
coupled to two optical modes both at Alice's and Bob's sites. The
mechanical elements are the remote modes, while the optical modes
are sent and shared with Charlie. Indeed, the goal of the protocol
is the creation and certification of entanglement without any
direct measurement on the mechanical elements, since quantum-limited measurement on mechanical
modes maybe highly nontrivial~\cite{Paternostro2010,DeChiara2011}. Therefore, it is
necessary to exploit the two output optical modes as Bell and
certifying modes (cf. Fig.~\ref{fig:scheme}(c)). This could be
done by driving a single cavity mode, and then extracting two
independent output optical modes by suitably filtering the
outgoing field as in~\cite{Genes2008}. However, it is more
efficient to drive \emph{two different} cavity modes and filtering
one output mode \cite{Giovannetti2001a,Genes2009} for each driven mode, and we
shall consider this latter situation from now on. The two filtered
optical modes are sent to Charlie for performing the Bell
measurement and the certifying process. The latter is only a
series of homodyne measurements, which will be carried out on the
optical modes only.

\subsection{The Hamiltonian}
The optomechanical system
is driven by two lasers which are appropriately detuned from the corresponding cavity mode.
Thus, the Hamiltonian of the system is described by $\hat{H}_{\mathrm{sys}}=\hat{H}_{\rmO} +\hat{H}_{\rmM} +\hat{H}_{\rmO\rmM} +\hat{H}_{\rmL}$, where
\begin{equation}
\hat{H}_{\rmO}=\hbar\omega_{b}\hat{a}_{b}^{\dagger}\hat{a}_{b} +\hbar\omega_{c}\hat{a}_{c}^{\dagger}\hat{a}_{c}
\end{equation}
describes two different modes of the optical cavity with frequencies $\omega_{b}$ and $\omega_{c}$ and whose annihilation operators satisfy the usual bosonic commutation relations $[\hat{a}_{k},\hat{a}_{k'}]=[\hat{a}_{k}^{\dagger},\hat{a}_{k'}^{\dagger}]=0$ and $[\hat{a}_{k},\hat{a}_{k'}^{\dagger}]=\delta_{kk'}$ with $k,k'=b,c$.
The mechanical element is described by
\begin{equation}
\hat{H}_{\rmM}=\frac{\hbar\omega_{\rmM}}{2}\left(\hat{p}^{2}+\hat{q}^{2}\right),
\end{equation}
which corresponds to a mechanical oscillator with mass $m$ and resonance frequency $\omega_{\rmM}$.
This means assuming that the cavity modes interact only with one resonant mode of the mechanical part of the system, which is justified when the detection bandwidth is chosen so that it includes only a single, isolated, mechanical resonance and mode--mode coupling is negligible \cite{Genes2008a}.
In the above mechanical Hamiltonian, $\hat{p}$ and $\hat{q}$ are the dimensionless momentum and position of the micro-mechanical oscillator, respectively, such that $[\hat{q},\hat{p}]=i$.
The optomechanical interaction is described by
\begin{equation}
\hat{H}_{\rmO\rmM}=-\hbar\hat{q}(G_{0,b}\hat{a}_{b}^{\dagger}\hat{a}_{b} +G_{0,c}\hat{a}_{c}^{\dagger}\hat{a}_{c}),
\end{equation}
where $G_{0,k}$ ($k=b,c$) are the single-photon optomechanical coupling constant. In the paradigmatic case of an optomechanical system formed by a Fabry-Perot cavity with a micromechanical mirror this coupling constants can be written in terms of the cavity length $L$ as~\cite{Genes2009,Aspelmeyer2012}
\begin{equation}
G_{0,k}\equiv \frac{\omega_{k}}{L}\sqrt{\frac{\hbar}{m\omega_{\rmM}}}.
\end{equation}
Finally, the laser driving is described by
\begin{equation}
\hat{H}_{\rmL}=i\hbar\big(E_{b}\hat{a}_{b}^{\dagger}e^{-i\omega_{\rmL,b}}+E_{c}\hat{a}_{c}^{\dagger}e^{-i\omega_{\rmL,c}}\big) +h.c. ~,
\end{equation}
where $|E_{k}|\equiv \sqrt{2\kappa_{k}P_{\rmL,k}/\hbar\omega_{\rmL,k}}$ is the driving rate of the cavity modes.
Here $P_{\rmL,k}$ is the laser input power and $\omega_{\rmL,k}$ its frequency, while $\kappa_{k}$ is the decay rate of the $k$th cavity mode.

\subsection{Quantum Langevin equations}
We use a quantum Langevin equation (QLE) approach to study the quantum dynamics of the optomechanical system at each site.
The QLE can be derived from the full Hamiltonian of the system, i.e., by adding the Hamiltonian of the mechanical and optical reservoirs and their interaction with the system to $\hat{H}_{\mathrm{sys}}$ yielding, in a frame rotating at the frequencies of the two lasers \cite{Giovannetti2001},
\begin{subequations}
\begin{eqnarray}
\dot{\hat{q}} &=& \omega_{\rmM}\hat{p} \\
\dot{\hat{p}} &=& -\omega_{\rmM}\hat{q} -\gamma_{\rmM}\hat{p} +G_{0,b}\hat{a}_{b}^{\dagger}\hat{a}_{b} +G_{0,c}\hat{a}_{c}^{\dagger}\hat{a}_{c} +\hat{\xi} \\
\dot{\hat{a}}_{b} &=& -[\kappa_{b}+i(\Delta_{0,b}-G_{0,b}\hat{q})]\hat{a}_{b} +E_{b} +\sqrt{2\kappa_{b}}~\hat{a}_{b}^{\rmin}\\
\dot{\hat{a}}_{c} &=& -[\kappa_{c}+i(\Delta_{0,c}-G_{0,c}\hat{q})]\hat{a}_{c} +E_{c} +\sqrt{2\kappa_{c}}~\hat{a}_{c}^{\rmin}
\end{eqnarray}
\end{subequations}
where $\Delta_{0,k}\equiv \omega_{k}-\omega_{\rmL,k}$ is the detuning of the laser frequency with respect to the cavity modes.
The mechanical noise operator $\hat{\xi}$ describes the zero-mean thermal noise, with correlation function
\begin{equation}
\langle\hat{\xi}(t)\hat{\xi}(t')\rangle=\frac{\gamma_{\rmM}}{\omega_{\rmM}} \int\frac{d\omega}{2\pi}\omega e^{-i\omega (t-t')}\big[1+\coth(\frac{\hbar\omega}{2k_{\mathrm{B}}T})\big],
\end{equation}
where $\gamma_{\rmM}$ is the damping rate, $k_{\mathrm{B}}$ is the Boltzmann constant, and $T$ is temperature of the mechanical reservoir.
The only non-vanishing correlation function of the noise operators acting on the optical modes due to the vacuum fluctuations are
\begin{equation}
\langle\hat{a}_{k}(t)\hat{a}_{k'}^{\dagger}(t)\rangle =\delta_{kk'}\delta(t-t').
\end{equation}
In the present proposal, non-local entanglement between the two non-interacting mechanical resonators at Alice and Bob site is created by swapping an initially present optomechanical entanglement between the mechanical mode and the Bell output optical mode. This latter entanglement is known to be strong and robust in the case of strong optomechanical coupling \cite{Vitali2007,Hofer2011,Abdi2011}, and a straightforward way to enter this regime~\cite{Groblacher2009,Teufel2011} is to intensely drive the optical modes and to consider the linearized quantum fluctuations around the resulting classical steady state.
By assuming high intensity intracavity fields one approximates the cavity mode operators as a steady state coherent field with large amplitude and quantum fluctuations around it.
Therefore, for every operator $\hat{o}$ one can write $\hat{o}=o_{\rms}+\delta\hat{o}$ and get the following classical steady state values
\begin{subequations}
\begin{eqnarray}
p_{\rms} &=& 0, \\
q_{\rms} &=& \frac{1}{\omega_{\rmM}}\big(G_{0,b}|a_{\rms,b}|^{2}+G_{0,c}|a_{\rms,c}|^{2}\big), \\
a_{\rms,k} &=& \frac{E_{k}}{\kappa_{k}+i\Delta_{k}}, ~~~(k=b,c),
\end{eqnarray}
\end{subequations}
where the effective detuning are defined as $\Delta_{k}\equiv \Delta_{0,k} -G_{0,k}q_{\rms}$.

The linearized dynamics of the small quantum fluctuations of the optomechanical system can be described in compact form in terms of the vector of fluctuations $\hat{\mathbf{u}}\equiv [\delta\hat{q},\delta\hat{p},\delta\hat{x}_{b},\delta\hat{y}_{b},\delta\hat{x}_{c},\delta\hat{y}_{c}]^{\trp}$ as
\begin{equation}
\dot{\hat{\mathbf{u}}}=\K\hat{\mathbf{u}}+\hat{\mathbf{n}},
\end{equation}
where $\hat{\mathbf{n}}\equiv [0,\hat{\xi},\sqrt{2\kappa_{b}}\hat{x}_{b}^{\rmin},\sqrt{2\kappa_{b}}\hat{y}_{b}^{\rmin},\sqrt{2\kappa_{c}}\hat{x}_{c}^{\rmin},\sqrt{2\kappa_{c}}\hat{y}_{c}^{\rmin}]^{\trp}$ is the noise vector, and $\K$ is the matrix of coefficients, given by
\begin{equation}
\K \equiv \left[
\begin{array}{cccccc}
0 & \omega_{\rmM} & 0 & 0 & 0 & 0\\
-\omega_{\rmM} & -\gamma_{\rmM} & G_{b} & 0 &
G_{c} & 0\\
0 & 0 & -\kappa_{b} & \Delta_{b} & 0 & 0\\
G_{b} & 0 & -\Delta_{b} & -\kappa_{b} & 0 & 0\\
0 & 0 & 0 & 0 & -\kappa_{c} & \Delta_{c}\\
G_{c} & 0 & 0 & 0 & -\Delta_{c} & -\kappa_{c}%
\end{array}
\right],
\end{equation}
where $G_{k}\equiv \sqrt{2}G_{0,k}a_{\rms,k}$ are the effective optomechanical couplings which can be made large and tunable by varying the stationary intracavity amplitudes $a_{\rms,k}$.

The steady state of the tripartite optomechanical system exists and it is stable if all the eigenvalues of the drift matrix $\K$ have negative real parts.
The parameter region under which stability occurs can be obtained from the Routh--Hurwitz criterion \cite{Ogata2010}, but the inequalities that come out are quite involved. However, the present bichromatically driven system has a regime in which the system is always stable, achieved when $G_{b}=G_{c}$ and $\Delta_{b}=-\Delta_{c}$, where there is a balance between a stabilizing ``cooling'' cavity mode with positive detuning and a ``heating'' cavity mode with negative detuning.
Ref.~\cite{Genes2009} has shown that this bichromatically driven system in this regime provides a robust and significative optomechanical entanglement and we assume to operate in such a regime for a possible implementation of the proposed entanglement swapping protocol.

\subsection{Optomechanical entanglement of output modes}
Charlie performs his Bell and certifying measurements on the optical modes at the output of the optomechanical cavities, which can always be optimized with filters which, if appropriately chosen, may lead to a significative increase of the entanglement with respect to their intracavity counterpart~\cite{Genes2008}.
The effective, filtered output modes are defined by the following bosonic annihilation operators
\begin{equation}
\hat{a}_{k}^{\rmout}(t)=\int_{t_{0}}^{t}h_{k}(t-s)\big[\sqrt{2\kappa_{k}} \delta\hat{a}_{k}(t)-\hat{a}_{k}^{\rmin}(t)\big]ds
\end{equation}
where $h_{k}(t)$ is a causal filter function defining the output modes~\cite{Genes2008}.
In fact, $\hat{a}_{k}^{\rmout}$ is a standard photon annihilation operator, implying the normalization condition $\int|h_{k}(t)|^{2}dt=1$.
A simple choice is
\begin{equation}
h_{k}(t)=\sqrt{\frac{2}{\tau_k}}\Theta(t)\exp\big[-(\frac{1}{\tau_k}+i\Omega_{k})t\big],
\end{equation}
where $\Theta(t)$ is the Heaviside step function, $1/\tau_k$ is the bandwidth of the filter, and $\Omega_{k}$ is the central frequency (measured with respect to the frequency of the corresponding driving field).

The stationary entanglement in the tripartite Gaussian state of the selected output optical modes and the mechanical resonator is determined by its $6\times 6$ CM
\begin{equation}
\V_{ij}^{\rmout}=\frac{1}{2}\big\langle
\hat{u}_{i}^{\rmout}(\infty)\hat{u}_{j}^{\rmout}(\infty)
+\hat{u}_{j}^{\rmout}(\infty)\hat{u}_{i}^{\rmout}(\infty)\big\rangle
,
\end{equation}
where $\hat{\mathbf{u}}^{\rmout} \equiv [\delta\hat{q},\delta\hat{p},\hat{x}_{b}^{\rmout},\hat{y}_{b}^{\rmout},\hat{x}_{c}^{\rmout},\hat{y}_{c}^{\rmout}]^{\trp}$ is the vector formed by the output field quadratures and by the mechanical operators.
This output CM can be expressed in terms of a frequency integral as~\cite{Genes2008,Genes2009}
\begin{align}
\V^{\rmout}=\int d\omega\tilde{\T}(\omega)\big[\tilde{\N}(\omega)+\cP_{\rmout}\big]\Q(\omega) \nonumber \\
\times\big[\tilde{\N}(\omega)^{\dagger}+\cP_{\rmout}\big]\tilde{\T}(\omega)^{\dagger},
\end{align}
where $\tilde{\N}(\omega)\equiv(i\omega\I +\K)^{-1}$ and $\tilde{\T}(\omega)$ is the Fourier transform of
\begin{widetext}
\begin{equation}
\T(t)=\left[
\begin{array}{cccccc}
\delta(t) & 0 & 0 & 0 & 0 & 0 \\
0 & \delta(t) & 0 & 0 & 0 & 0 \\
0 & 0 & \sqrt{2\kappa_{b}}h_{b}^{\Re}(t) & -\sqrt{2\kappa_{b}}h_{b}^{\Im}(t) & 0 & 0 \\
0 & 0 & \sqrt{2\kappa_{b}}h_{b}^{\Im}(t) & \sqrt{2\kappa_{b}}h_{b}^{\Re}(t) & 0 & 0 \\
0 & 0 & 0 & 0 & \sqrt{2\kappa_{c}}h_{c}^{\Re}(t) & -\sqrt{2\kappa_{c}}h_{c}^{\Im}(t) \\
0 & 0 & 0 & 0 & \sqrt{2\kappa_{c}}h_{c}^{\Im}(t) & \sqrt{2\kappa_{c}}h_{c}^{\Re}(t)
\end{array}
\right].
\end{equation}
\end{widetext}
$\cP\equiv\mathrm{diag}[0,0,1/2\kappa_{b},1/2\kappa_{b},1/2\kappa_{c},1/2\kappa_{c}]$ is the projector onto the optical quadratures, while $\Q(\omega)$ is the diffusion matrix of the system, given by
\begin{equation}
\Q(\omega)=\mathrm{diag}\Big[0,\frac{\gamma_{\rmM}}{\omega_{\rmM}}\omega\coth\big(\frac{\hbar\omega}{2k_{\rmB}T}\big),\kappa_{b},\kappa_{b},\kappa_{c},\kappa_{c}\Big].
\end{equation}
Using the CM one can analyze the bipartite entanglement within the three different bipartitions of the system when one of the three modes is traced out, and also the tripartite entanglement.

\subsection{ Entanglement of the micromechanical resonators by entanglement swapping }
An initially present optomechanical entanglement between the mechanical resonator and an output cavity mode (in each tripartite system) can be swapped into an entanglement between the two remote mechanical oscillators by means of the Bell measurement on the two optical modes.
Furthermore, such an entanglement can be locally verified and certified by Charlie when there is a nonzero entanglement between the two optical certifying output fields, one from Alice and the other from Bob. From the discussion of Sec.~III, the two above conditions are achieved when the tripartite optomechanical systems at each site is initially in a state satisfying the certifying condition of Eq.~(\ref{cert}), involving only purities. Therefore, we have to determine an experimentally achievable parameter set in which such conditions are satisfied so that the proposed generalized swapping protocol can be successfully implemented.

Still restricting to the symmetric case of initially identical
states at Alice and Bob sites, one has the following
classification of tripartite optomechanical states:
\begin{eqnarray*}
Class~1)&~certifiable:~~~~~~~~~&\mu_{\rmR\rmB}>\mu_{\rmB\rmC}>\mu_{\rmB} \\
Class~2)&~not~certifiable:~~~~&\mu_{\rmR\rmB}>\mu_{\rmB} ~\& ~\mu_{\rmB\rmC}<\mu_{\rmB} \\
Class~3)&~wrong~swapping:~~&\Big\{
\begin{array}{l}
\mu_{\rmB\rmC}>\mu_{\rmR\rmB}>\mu_{\rmB} \\
\mu_{\rmB\rmC}>\mu_{\rmB} ~\& ~\mu_{\rmR\rmB}<\mu_{\rmB}
\end{array} \\
Class~4)&~no~swapping:~~~~~~~&\mu_{\rmR\rmB}<\mu_{\rmB} ~\&
~\mu_{\rmB\rmC}<\mu_{\rmB}
\end{eqnarray*}
The first case is the desired class of tripartite \textit{certifying} states~\cite{Abdi2012} which guarantees a successful implementation of the protocol.
In the second case, the two remote mechanical resonators are entangled after the protocol, but there is no entanglement between the certifying optical modes.
Therefore, the success of entanglement swapping cannot be locally certified.
In the third case, which we call ``wrong swapping'' the Bell and certifying modes are more entangled than the remote and Bell modes; in this case one has an entangled pair of certifying modes, but this entanglement is greater than the value of the mechanical--mechanical entanglement which can be either zero or nonzero, and therefore one cannot say anything certain about the entanglement between the remote modes.
In this case in fact, instead of having most entangled mechanical resonators, one gets two optical modes with higher entanglement.
The fourth case is the worst situation when no swapping occurs because we do not have the necessary optomechanical entanglement to start with.

The desired certifying condition of Eq.~(\ref{cert}) is satisfied if we appropriately choose the detuning and filter the output modes.
In fact, we have found that the mechanical--mechanical entanglement is larger when we drive the cavity Bell mode with a blue-detuned laser ($\Delta_{b} < 0$) and the certifying mode by a red-detuned laser ($\Delta_{c} > 0$).
Indeed, it is shown in Ref.~\citep{Genes2009} that in this case the remote--Bell optomechanical entanglement is larger and both required conditions of large mechanical--mechanical entanglement and smaller certifying entanglement are easier to achieve. Let us verify this by considering an optomechanical system with state-of-the-art parameter values.
The system is composed of a Fabry-P\'erot cavity with $L=1~\mathrm{mm}$ length, whose movable mirror has an effective mass $m=10~\mathrm{ng}$, resonance frequency of $\omega_{\rmM}/2\pi=10~\mathrm{MHz}$, quality factor $Q_{\rmM}=\gamma_{\rmM}/\omega_{\rmM}=10^5$, and coupled to a reservoir at temperature $T=0.4~\mathrm{K}$.
We consider two lasers driving two adjacent cavity modes with wavelengths $\lambda_b=810.045~\mathrm{nm}$ and $\lambda_c=810.373~\mathrm{nm}$ and with the above-mentioned choice of opposite detunings, $\Delta_b=-\Delta_c=-\omega_{\rmM}$.
Moreover the output optical modes corresponding to the Bell modes are filtered in order to be centered around the Stokes sideband, while the certifying modes are centered around the anti-Stokes sideband, i.e., we have $\Omega_b=-\Omega_c=-\omega_{\rmM}$. We now study the properties of the initial tripartite Gaussian state with the above parameter choice, as a function of the remaining parameters, that is, the cavity bandwidths $\kappa_k$, the input powers $P_k$, and the bandwidth of the filtered output modes, $1/\tau_k$, $k=b,c$.
\begin{figure}[t]
\includegraphics[width=\columnwidth]{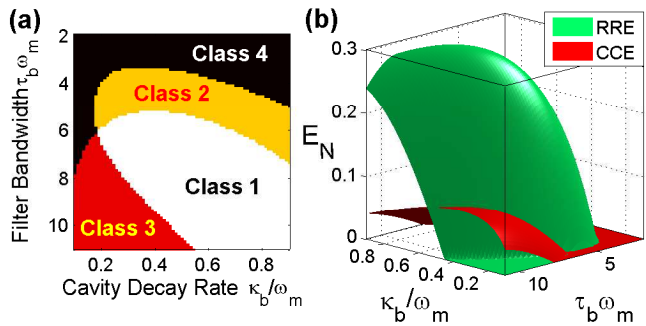}
\caption{(Color online) (a) Classification of the input tripartite states, and (b) value of the $E_N$ for the entanglement between the remote modes `RRE' and the certifying modes `CCE' as a function of the cavity decay rates and filtering bandwidths. The system parameters are: input powers $P_b=4~\mathrm{mW}$ and $P_c=4.5~\mathrm{mW}$, detuning of the lasers $\Delta_b=-\Delta_c=-\omega_{\rmM}$, filtering inverse bandwidths are chosen so that $\tau_c=\tau_b/6$, and the decay rate are chosen to be equal $\kappa_c=\kappa_b$. See the text for the other parameters.}\label{fig:kaptau}
\end{figure}

Fig.~\ref{fig:kaptau}(a) shows the class of the initial Gaussian
tripartite state at fixed input powers $P_b=4~\mathrm{mW}$,
$P_c=4.5~\mathrm{mW}$, in a chosen interval of cavity bandwidths
(here assumed to be equal $\kappa_b = \kappa_c$) and of inverse
output bandwidths (here chosen so that $\tau_c=\tau_b/6$). The
white region corresponds to the desired class 1 of certifying
states, leading to a successful entanglement swapping certifiable
with local measurements. Fig.~\ref{fig:kaptau}(b) refers to the
same parameter region and describes the ``output'' of the
protocol. In fact, it shows the logarithmic negativity $E_N$ of
the mechanical--mechanical entanglement (the green surface named
as `RRE'), and of the certifying optical modes (the red surface
named as `CCE'). This latter figure shows that a log-negativity
RRE of $E_N \simeq 0.3$ for the remote modes can be certified by
$E_N \simeq 0.05$ for the certifying modes.

\begin{figure}
\includegraphics[width=\columnwidth]{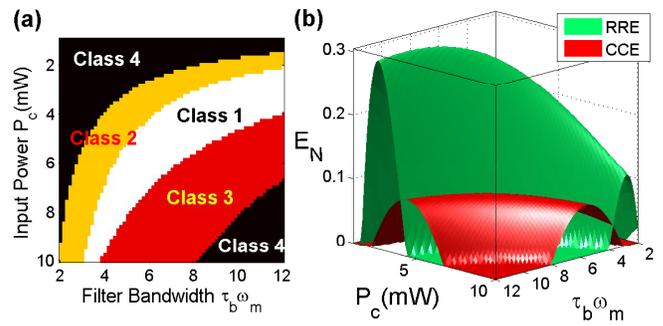}
\caption{(Color online)  (a) Classification of the input tripartite states, and (b) value of the $E_N$ for the entanglement between the remote modes `RRE' and the certifying modes `CCE' as a function of laser power and filtering bandwidths. The cavity decay rates are fixed, $\kappa_b=\kappa_c=0.5\omega_{\rmM}$, detuning of the lasers are $\Delta_b=-\Delta_c=-\omega_{\rmM}$. The filtering inverse bandwidth are chosen so that $\tau_c=\tau_b/5$, while the laser powers are chosen so that $P_c-P_b=0.5~\mathrm{mW}$. See the text for the other parameters.}\label{fig:powtau}
\end{figure}

Then, Fig.~\ref{fig:powtau} shows the class of the initial
Gaussian tripartite state and the protocol output as a function of
the input powers and filtering bandwidths (now assuming
$\tau_c=\tau_b/5$), at fixed and identical cavity bandwidths
$\kappa_b=\kappa_c=0.5\omega_{\rmM}$. In this case, the desired
certifying state region of class 1 is the white strip shown in
Fig.~\ref{fig:powtau}(a), while Fig.~~\ref{fig:powtau}(b) shows
again $E_N$ of the mechanical--mechanical entanglement (green
surface `RRE'), and of the certifying optical modes (red surface
`CCE'). Fig.~~\ref{fig:powtau}(b) indicates that a remote
entanglement of $E_N \simeq 0.2$ can be certified by an
entanglement between the two certifying optical modes $E_N \simeq
0.1$.

%
%
\section{Conclusion}
We have described in detail an extension of the entanglement
swapping protocol which can be applied to an appropriate class of
tripartite states. This protocol allows to swap an initially
available entanglement to two sites which have never interacted
and to \emph{certify} it by measuring locally the entanglement
between two ancillary modes at the same site where the Bell
measurement is carried out. We determine and characterize the
class of certifying states in the case of tripartite Gaussian CV
states, showing that they can be fully identified in terms of
local and bipartite purities~\cite{Abdi2012}.

We have then discussed the application of the proposed swapping
protocol with local certification to identical tripartite Gaussian
states of two optomechanical systems. The protocol is applied to
generate entanglement between two mechanical resonators at two
remote sites, using two output optical cavity modes from each site
for carrying out both the Bell measurement, for swapping the
entanglement, and the homodyne measurements for certifying the
success of the protocol. In this work we considered detections
performed on optical modes but our analysis could be extended to
other types of systems, for instance to the microwave modes of a
modified coplanar waveguide \cite{Barzanjeh2012}.
%
%
\section*{Acknowledgments}

This work has been supported by the European Commission (ITN-Marie
Curie project cQOM, FET-Open Project iQUOEMS), MIUR (PRIN
2010-2011), EPSRC (through HIPERCOM, EP/J00796X/1 and qDATA,
EP/L011298/1) and The Leverhulme Trust.

%
\bibliography{swapping}

\end{document}